\documentclass{elsart}

\usepackage{graphicx}
\usepackage[english]{babel}
\usepackage{natbib}
\usepackage{amssymb}
\usepackage{hyperref}

\graphicspath{{figures/}}

\begin{document}

\begin{frontmatter}

\title{Segment Boundary Detection via Class Entropy Measurements in Connectionist Phoneme Recognition\thanksref{titlethanks}.
}
\thanks[titlethanks]{This research was carried out at the Centre for Speech
    Technology supported by Vinnova (The Swedish Agency for Innovation
    Systems), KTH and participating Swedish companies and organisations.
}

\author{Giampiero Salvi}
\ead{giampi@kth.se}
\ead[url]{https://www.kth.se/profile/giampi}
\address{KTH,\\
  School of Computer Science and Communication, \\
  Speech, Music and Hearing,\\
  Stockholm, Sweden}

\begin{abstract}
This article\footnote{Please cite as:\\Giampiero Salvi, ``Segment Boundary Detection via Class Entropy Measurements in Connectionist Phoneme Recognition'', Speech Communication Volume 48, Issue 12, December 2006, Pages 1666-1676. \url{https://doi.org/10.1016/j.specom.2006.07.009}}  investigates the possibility to use the class entropy of
the output of a connectionist phoneme recogniser to predict time
boundaries between phonetic classes. The rationale is that the value
of the entropy should increase in proximity of a transition between
two segments that are well modelled (known) by the recognition
network since it is a measure of uncertainty. The advantage of this
measure is its simplicity as the posterior probabilities of each class
are available in connectionist phoneme recognition.

The entropy and a number of measures based on differentiation of the
entropy are used in isolation and in combination. The decision methods
for predicting the boundaries range from simple thresholds to neural
network based procedure.

The different methods are compared with respect to their precision,
measured in terms of the ratio between the number $C$ of predicted
boundaries within 10 or 20 msec of the reference and the total number
of predicted boundaries, and recall, measured as the ratio between $C$
and the total number of reference boundaries. 
\end{abstract}

\begin{keyword}
boundary detection \sep entropy \sep connectionist phoneme recognition
\PACS 43.72.Bs \sep 43.72.Ne \sep 43.60.Np
\end{keyword}

\end{frontmatter}

\section{Introduction}
In a number of speech analysis and recognition applications it is
essential to estimate the boundaries between phonetic segments. In
landmark \citep{gs:Liu1996} and segment based
\citep{gs:Glass2003,gs:OstendorfAndVeilleux1994} recognition this
information is used to find suitable locations in the speech signal
where to measure the features that constitute an acoustic observation
for further analysis.

In the Synface system \citep{gs:KarlssonEtAl2003} a phoneme recogniser
based on a hybrid of neural networks and hidden Markov models is used
to derive the lip movements of an avatar \citep{gs:Beskow2004}. The
avatar is used to provide lip reading support for hearing
impaired people communicating via telephone.
In this application both the recognition accuracy and the precision in
predicting the segment boundaries are important to synchronise the lip
movements to the acoustic signal.

The task of detecting phonetic boundaries is related, but not
equivalent, to that of aligning the speech signal to a reference
transcription \citep{gs:Hosom2002}. In alignment, the number of
phonetic boundaries is determined by a reference transcription, in
detection it is based exclusively on measures obtained from the speech
signal, adding a degree of freedom to the problem.

The use of connectionist models for phoneme classification
suggests the use of entropy measures based on the posterior probabilities
of the phonetic classes given an observation. These are calculated by
the neural network for each speech frame. The entropy of the posterior
probabilities has been shown to be a good confidence measure on the
frame-by-frame classification results \citep{gs:WilliamsAndRenals1999, gs:Salvi2006b}.

Here we focus on the time variation of the entropy and observe that,
as a measure of uncertainty, it should present local maxima at the
transitions between consecutive segments that are well modelled by the
recognition network. This can be explained by noting that, in
proximity to a boundary, a number of overlapping frames, and therefore
the resulting acoustic features vectors, contain characteristics of
both phonetic classes. The recognition network is likely to estimate,
in these cases, class probabilities with higher uncertainty.

In \cite{gs:Salvi2006} it was shown that the value of the entropy at
frame $n$ and its first and second derivatives are dependent to the
distance of the frame from a phonetic boundary. In this paper the
idea is further developed and differential measures of the entropy are
used as predictors of phonetic boundaries.

The different methods are compared counting the number $C$ of predicted
boundaries that fall within a number of milliseconds from the
reference boundaries. Two measures of performance are considered:
\emph{precision} relates $C$ to the total number of predicted
boundaries, while \emph{recall} relates $C$ to the total number of
reference boundaries.






The rest of the paper is organised as follows: Section~\ref{sec:framework}
describes the recogniser and the data used in the
experiments. Section~\ref{sec:method} explains the method and the measures
analysed in the experiments. Section~\ref{sec:results}
presents the results. Finally Section~\ref{sec:discussion} discusses
the results and Section~\ref{sec:conclusions} concludes the paper.

\section{The Framework} \label{sec:framework}
\subsection{The Recogniser}
The Synface recogniser is a hybrid of recurrent neural networks (RNNs),
and hidden Markov models (HMMs). The recogniser is described in
detail in \cite{gs:Salvi2003,gs:Salvi2006b}, here we mention only the
aspects that are relevant to this study.

The input layer of the RNN contains thirteen units that represent the
Mel frequency cepstral coefficients $C_0,...,C_{12}$.

The activities $a_i[n]$ of the output layer give an estimate of the
posterior probabilities $P(x_i|O[n])$ of each of the $N = N_\mathrm{p} +
N_\mathrm{n}$ acoustic classes $x_i$, given the observation
$O[n]$. The acoustic classes include $N_\mathrm{p}$ phonetic classes
and $N_\mathrm{n}$ noise and silence classes. The total number of acoustic
classes is 50 for Swedish, with 46 phonemes and 4 kinds of
noise/silence.

The posterior probabilities are fed into a Viterbi-like decoder where
the look-ahead length can be varied. The recognition network specified
by a Markov chain defines a loop of phonemes, where every phoneme
is represented by a three state left-to-right HMM.


\subsection{Data} \label{sec:data}
The recogniser was trained on the Swedish SpeechDat database
\citep{gs:Elenius2000}. The database has been divided into a training,
a validation and a test set, with 33062, 500 and 4150 utterances,
respectively. The validation set was used during training of the
recogniser to optimise a number of factors, e.g. the number of units
in the hidden layer, the type and number of connections between the
layers, and a number of parameters related to the training
procedure. Mel frequency cepstral coefficients were computed at every
10 msec.

The phonetic transcriptions used during training of the recogniser,
have been obtained with forced alignment. These transcriptions are
sufficiently accurate for the task of training the neural networks, but
cannot be safely adopted as a reference when the task is to evaluate
procedures for boundary detection. In order to obtain a more reliable
reference, a number of utterances, randomly chosen from the test
material, have been manually transcribed. These utterances are used in
the experiments described in the paper. Given the time consuming
process of manually transcribing speech material, only 80 test
utterances were used. In our opinion, the resulting total number of
boundaries (2816), constitutes a good statistical sample for the
experiments of this paper.

\begin{figure}
\centering
\includegraphics[scale=0.4]{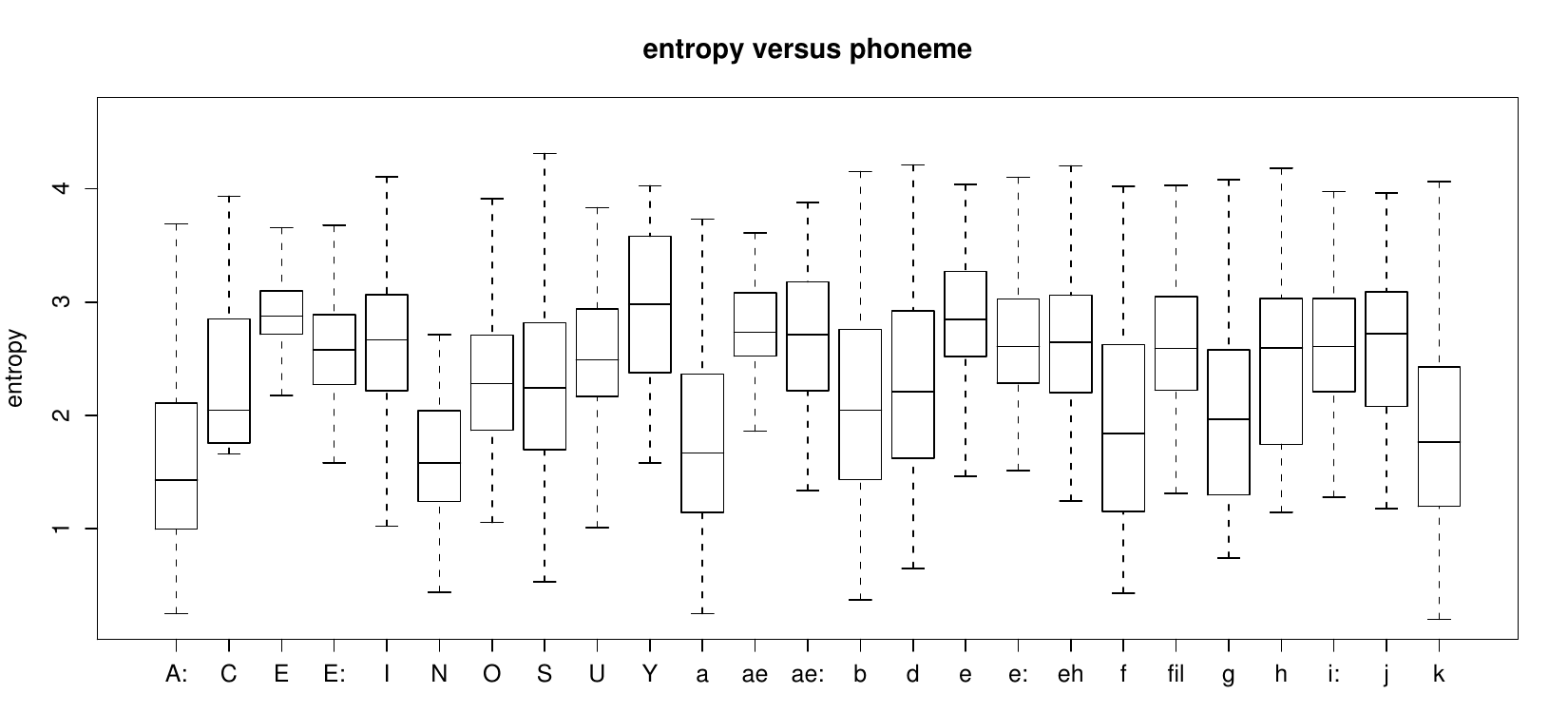}
\includegraphics[scale=0.4]{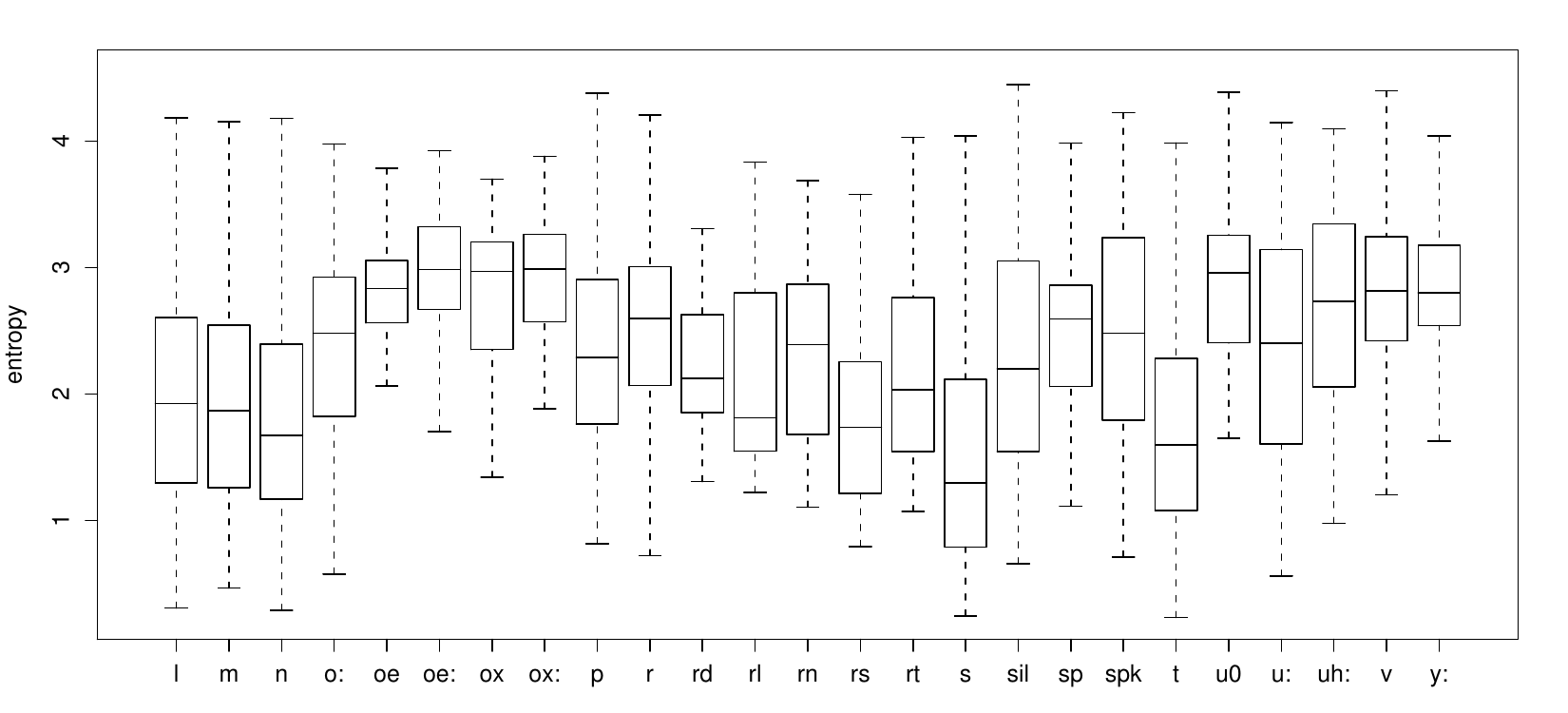}
\caption{Box plot of the frame entropy for each phonemic class. The
  maximum entropy is $\log_2 50 = 5.64$ bits. The SAMPA phonetic
  symbols \citep{gs:Wells1997} are used with a few
  exceptions \citep{gs:Lindberg2000}: \texttt{\} $\rightarrow$ uh, 2
  $\rightarrow$ ox, \{ $\rightarrow$ ae, 9 $\rightarrow$ oe, @
  $\rightarrow$ eh}}
\label{fig:entvssym}
\end{figure}
\section{Method} \label{sec:method}
As introduced in Section~\ref{sec:framework} the activities $a_i[n]$
of the output nodes of the phoneme recognition network give an
estimate of the posterior probabilities $P(x_i|O[n]),\ i\in[1,N]$ of
the acoustic class $x_i$ given the observation $O[n]$ at frame
$n$. The class entropy of the frame is defined as
\begin{equation}
 e[n] = -\sum_{i=1}^{N} a_i[n] \log_2 a_i[n] 
\label{eqn:entropy}
\end{equation}

Figure~\ref{fig:entvssym} shows the box plot of the entropy in
Equation~\ref{eqn:entropy} for each of the 50 phonetic classes
(including noise). The higher bound to the entropy is $\log_2 50 = 5.64$
bits.

The candidates for the boundaries are estimated by searching for local
maxima in the entropy function. This is a problem of
\emph{peak picking}, in relatively unusual conditions. Most of the
literature on peak picking
\citep[e.g.][]{gs:LiEtAl1995,gs:AndreevEtAl2003}, in fact, deals with
the problem of estimating the peaks (local maxima) of a process
affected by noise. Smoothing methods are employed to eliminate the
effect of the noise fluctuations, preserving the shape of the
``clean'' process. Once the smoothing is applied, the local maxima are
obtained by simply imposing that the first derivative of the measure,
with respect to time, is nearly 0, when the second derivative is
negative. This is possible when the function of time is sampled at a
sufficiently high rate compared with its time fluctuations.

In our case, the entropy is computed on a number of probability
estimators that, in turn, operate on a frame of speech samples. The
entropy is thus a smooth function of time, and the effects of
additive noise are limited. However, the main problem in our
application is that this function is sampled at a low rate, compared
to its fluctuations. One consequence is that the first derivative
looses its property of vanishing in proximity to a maximum (or minimum
or inflection point) and cannot be used in isolation to determine the
positions of the maxima.

\begin{figure}
\centering
\includegraphics[scale=0.4]{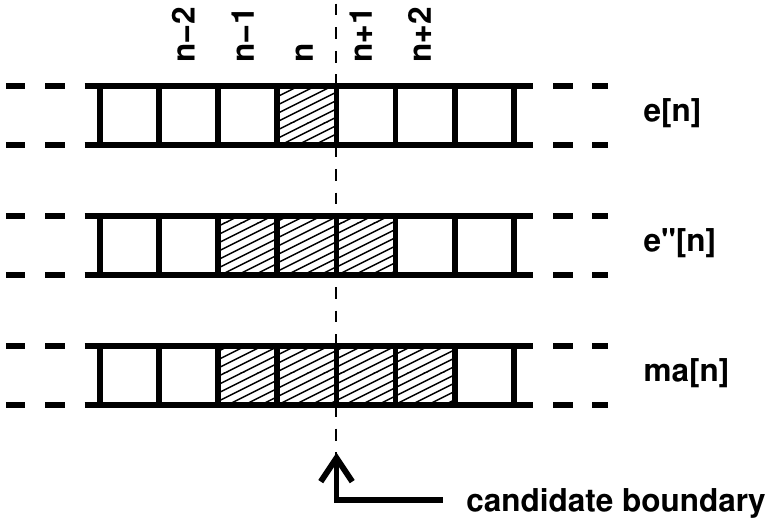}
\caption{Illustration of the time span of each measure with respect to
the candidate boundary} \label{fig:methodscheme}
\end{figure}

A number of methods have been investigated in order to find the local
maxima in the entropy function. The simplest method, referred to as
$\mathbf{e[n]}$ in the following, applies a threshold on the entropy
function.
As illustrated in Figure~\ref{fig:entvssym}, the entropy is not only
dependent on the proximity to a boundary as postulated up to now, but
also on the phonetic segment. This limits the effectiveness of this
simple method, as it is hard to find a single value for the threshold
that would work in general.

A solution to this is the use of differential measures. As already
mentioned, the first derivative of the entropy in these conditions
bears limited information about the position of the boundaries.
The second derivative defined as 
\[
 e^{\prime\prime}[n] = e^\prime[n+1]-e^\prime[n] = e[n-1]-2e[n]+e[n+1]
\]
on the contrary, is a good indication of local maxima, as low values
correspond to regions where the function $e[n]$ is concave, and, given
the low sampling rate, these regions are likely to contain local maxima.

\newcommand{\nmo}{\ensuremath{[n\!-\!1]}}
\newcommand{\npo}{\ensuremath{[n\!+\!1]}}
\newcommand{\npt}{\ensuremath{[n\!+\!2]}}
\newcommand{\grc}{\ensuremath{\stackrel{\star}{>}}}

As illustrated in Figure~\ref{fig:methodscheme} each sample of the
entropy function refers to a segment (frame), while the boundaries
are between two adjacent frames. It is therefore desirable to
base the decision on whether there is a boundary between frame $n$ and
frame $n+1$ on a measure that is symmetric with respect to these two
frames. One way of doing this is to average the second derivative over
two adjacent frames (moving average of the second order) that, besides
a constant factor, is:
\[
\mbox{ma}[n] = e^{\prime\prime}[n]+e^{\prime\prime}[n+1] =
e[n-1]-e[n]-e[n+1]+e[n+2]
\]

\begin{table}
\caption{Summary of the boundary detection methods, The difference
  between the decision indicated by the symbols $>$ and $\grc$ is
  explained in Section~\ref{sec:decision} and
  Figure~\ref{fig:decisionmethod}.} \label{tab:methods}
\footnotesize
\begin{tabular}{llp{0.3\textwidth}c} \hline\hline
method                         & measure           & definition & decision \\ \hline
$\mathbf{e[n]}$                & entropy           & $-\sum a_i[n]\log_2 a_i[n]$ & $e[n]\grc\mbox{th}$ \\
$\mathbf{e^\prime[n]}$         & first derivative  & $ e[n]-e\nmo$ & none \\
$\mathbf{e^{\prime\prime}[n]}$ & second derivative & $ e^\prime\npo-e^\prime[n] = e\nmo-2e[n]+e\npo$ & $-e^{\prime\prime}[n]\grc\mbox{th}$ \\
$\mathbf{\mbox{\bf ma}[n]} $   & moving average    & $ e^{\prime\prime}[n]+e^{\prime\prime}\npo = e\nmo-e[n]-e\npo+e\npt$ & $-\mbox{ma}[n]\grc\mbox{th}$ \\ \hline
method                              & measure & \multicolumn{2}{l}{decision} \\ \hline
$\mathbf{e[n]+e^{\prime\prime}[n]}$ & entropy + 2nd deriv. & \multicolumn{2}{l}{$ (e[n]> \mbox{th1})\ \ \cap\ \ (-e^{\prime\prime}[n] \grc\mbox{th2})$} \\
$\mathbf{e[n]+\mbox{\bf ma}[n]}$    & entropy + mov. av. & \multicolumn{2}{l}{$(e[n]> \mbox{th1})\ \ \cap\ \ (-\mbox{ma}[n] \grc\mbox{th2})$} \\
$\mbox{\bf nn}\mathbf{(e,e^\prime,e^{\prime\prime},\mbox{\bf ma})}$ & neural network & \multicolumn{2}{l}{$\mbox{nn}(e,e^\prime,e^{\prime\prime},\mbox{ma}) \grc \mbox{th}$} \\ \hline\hline
\end{tabular}
\end{table}

Table~\ref{tab:methods} gives a summary of the methods used in our
experiments. As shown in the table the boundaries were predicted by
setting a threshold on each of the above measures in isolation, or by
combining them. Two ways of combining the measures were
considered. The first way is motivated by the attempt to reduce the
insertions of boundaries obtained with the second derivative and with
the moving average methods. The absolute value of the entropy is used
in this case to eliminate those cases in which local maxima are
detected in regions of globally low entropy (spurious maxima).
The second combination method makes use of a time delayed neural
network with the regression task to estimate a function of proximity
to a boundary. In this case all the measures are given in input to the
network in order to let the training procedure find an optimal
combination. As the target values to the neural network should be
bounded between known limits (e.g. 0 and 1), it was necessary
to map the distance $d(f,b) \in [0,\infty)$ between frame $f$ and
boundary $b$ to the interval $(0,1)$. A convenient way of doing this
is by using the exponential function. The resulting measure, defined
as:
\begin{equation} \label{eqn:proximity}
 p(f,b) = e^{-d(f,b)}
\end{equation}
will be called proximity in the following.
A frame adjacent to a boundary has distance $d=0$ and proximity $p=1$.
As the distance $d\rightarrow\infty$ the proximity $p\rightarrow 0$.
The neural network is composed by an input, a hidden and an output
layer of respectively 4, 11 and 1 units. The layers are fully connected
with time-delayed arcs.
The estimate of the function $p(f,b)$ by the neural network will be
called $\mbox{\bf nn}\mathbf{(e,e^\prime,e^{\prime\prime},\mbox{\bf ma})}$
in the following.

\subsection{Decision Method}
\label{sec:decision}
The decision procedure indicated by the relation $m>\mbox{th}$ simply
finds the regions in which the measured $m$ exceeds the threshold
$\mbox{th}$.

\begin{figure}
\centering
\includegraphics[scale=0.4]{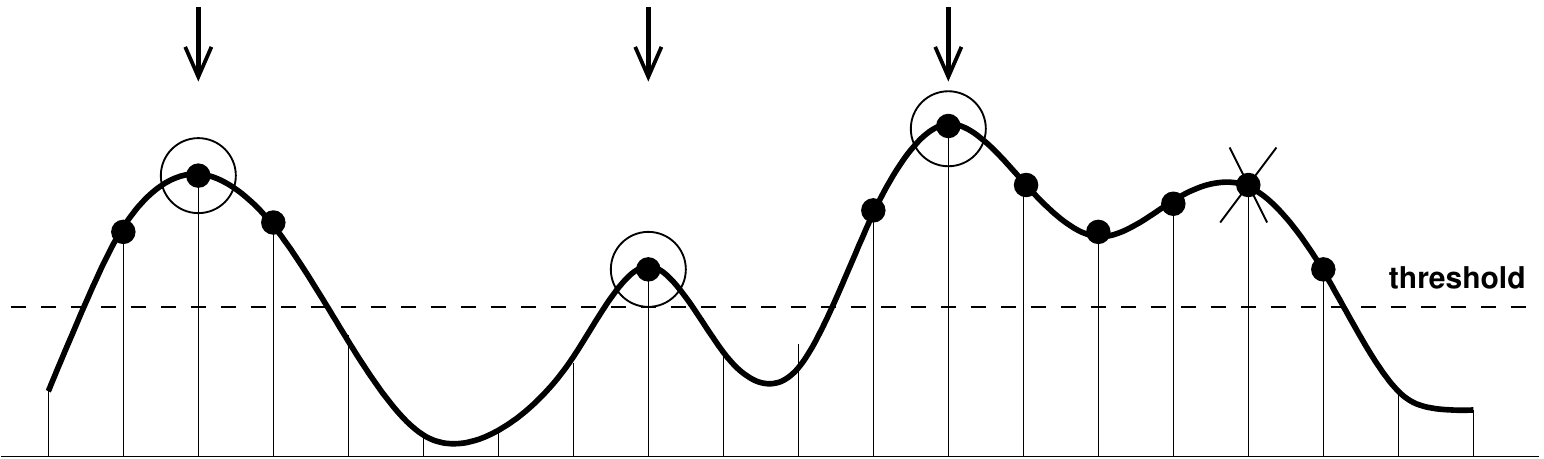}
\caption{Illustration of the $\grc$ decision method: given a measure
  $m$, one boundary candidate is selected for each region of
  contiguous frames for which $m>\mbox{th}$. The candidate is the
  global maximum for $m$ in that region. The vertical arrows indicate
  the resulting boundaries, the cross corresponds to a possible
  deletion.} \label{fig:decisionmethod}
\end{figure}

The decision procedure indicated by the relation $m\grc\mbox{th}$,
instead, picks only one boundary candidate for each region of contiguous
frames for which the measure exceeds the threshold. This avoids
detecting adjacent boundary candidates, but leads to boundary
deletions in case two or more local maxima exist in the same region as
illustrated in Figure~\ref{fig:decisionmethod}.

In case of combined measures, the regions in which the entropy $e[n]$
exceeds the threshold are first found with method $>$, then, within these
regions method $\grc$ is applied to the second measure
($e^{\prime\prime}[n]$ or $\mbox{ma}[n]$).

\subsection{Threshold}
The value of the threshold in the decisions above can be used to find
a trade-off between number of boundary insertions and deletions. The
range of suitable values for the thresholds depends on the measurements
they are applied to. In the following, all threshold values are relative
to the global statistics for each measured variable. Calling
$\mu_i$ and $\sigma_i$ respectively the mean and standard
deviation for the $i$th variable, the value for the corresponding
threshold ranges from $\mu_i-\sigma_i$ to
$\mu_i+2\sigma_i$ and will be indicated with relative values from -1.0
to 2.0.

When multiple measures are combined, an optimisation is performed to
determine the best combination of values.

\subsection{Baseline}
As the phoneme recognition network estimates the posterior
probabilities for each phonetic class given the observation (frame), its
output can be used directly as a classifier of each input frame in a
\emph{maximum a posteriori} sense. The phonetic boundaries could be
therefore detected by simply observing the changes in classification
from frame to frame. This method constitutes the baseline of our
investigation.

\subsection{Level of Chance}
\label{sec:levelofchance}
To determine the advantage in using any of the above methods, it is
interesting to measure the results obtained by randomly assigning $F$
boundaries to the $M$ possible positions, where $M+1$ is the number of
frames in the utterance. We call this measure \emph{level of chance}.

First we notice that, disregarding the effect of the beginning and end
 of each utterance, if we have $B$ non contiguous reference boundaries in $M$
 positions, the probability that a randomly chosen position $f$ is
 within one frame from any of the $B$ boundaries $b_i$ is
\[ \sum_{i=1}^{B} p(d(f,b_i)\leq 1) \approx \frac{3B}{M} \]
If we consider $F$ randomly selected positions and $B\ll M$ and $F\ll
M$, the above approximation is still valid, and the level of chance,
that is the ratio of positions $f_j$ within one frame of any reference
boundary is simply $\frac{3B}{M}$, independently of $F$. Similarly the
level of chance will be $\frac{5B}{M}$ if we consider boundaries with
at most 20 msec distance from the reference.
In the test material, the percentage of reference boundary per frame
is $B/M \simeq 0.114$ which gives a theoretical level of chance of
34.2\% for 10 msec and 57.0\% for 20 msec.

This value was verified with Monte Carlo simulations where the
detected boundaries were obtained by sampling without repetition the
available $M$ positions. The size of the sample $F$ was a predefined
ratio of the number $B$ of boundaries in the reference transcription
for each utterance. The ratio was varied in a way to simulate the
effect of varying the threshold (and thus the number of insertions and
deletions) in the methods described above. For each ratio, the
simulation was repeated 50 times over which the results were
averaged.

\subsection{Forced Alignment}
\label{sec:forcedalign}
Another reference for the performance of the methods considered in
this paper are the results obtained with forced alignment using the
recognition models, and the original transcriptions. It should be
clear in this case that the problem of detecting phoneme boundaries is
completely different from the problem of aligning an already known
sequence of phonemes to the acoustic observations. In the first case,
in fact, no \emph{a priori} information on the correct sequence of
phonemes (and as a consequence of the number of phonetic boundaries)
is given, introducing a number of degrees of freedom that make the
problem more challenging. The results are thus not directly
comparable, but the forced alignment results could be interpreted
as a higher bound to the boundary detection methods.

\subsection{Evaluation}
\label{sec:evaluation}
Two evaluation measures are used to compare the different methods. In
both cases the number of correctly detected boundaries $C$ is first
computed. A correctly detected boundary is defined as a boundary that
lies within a number of milliseconds from a reference boundary.

Two definitions of the time limit have been
used in this paper following the literature: \cite{gs:Liu1996} uses
10~msec as a limit of correctness, while \cite[e.g.][]{gs:Hosom2002}
uses 20~msec. Most of the analysis are performed with the 10 msec
limit, while the final results are presented for both limits.

Calling $D$ the total number of detected boundaries and $T$ the total
number of reference (transcription) boundaries, the two measures of
performance are: \emph{precision} $P$ computed as $C/D\%$, and
\emph{recall} $R$ computed as $C/T\%$.
Precision counts the percentage of correctly predicted boundaries with
respect of the total number of predicted boundaries.
Recall counts the percentage of correctly predicted boundaries with
respect of the total number of reference boundaries.

In the cases when an optimisation is performed, the above two measures
are combined in a single criterion that is defined as the Euclidean
distance between the point defined as $(P,R)$, and the
point (100\%,100\%), that is:
\begin{equation}
 \mbox{crit} = \sqrt{(P-100)^2+(R-100)^2}
 \label{eq:criterion}
\end{equation}
The optimisation tends to minimise this criterion. Note that this
criterion gives the same weight to precision and recall. In some
applications it may be more convenient to prioritise one of the two.

The procedure, described above, for computing the number of correctly
detected boundaries overlooks insertions. If two or more detected
boundaries are in the proximity of the same reference boundary, they
are all considered to be correct. This does not constitute a
limitation in most of the cases considered in this paper, as the
definition of the decision method $\grc$ prevents (or reduces)
detecting boundaries that are too close in time. The baseline method,
however, does not have such a constraint, and can in principle predict
a phonetic boundary for each time step. In this case, the number $C$
of correct boundaries could exceed the number of reference boundaries
$T$. This will affect both the measure of precision and recall that
can eventually exceed 100\%.

\section{Results}\label{sec:results}
\subsection{Example}
\begin{figure}
\centering
\includegraphics[width=\textwidth]{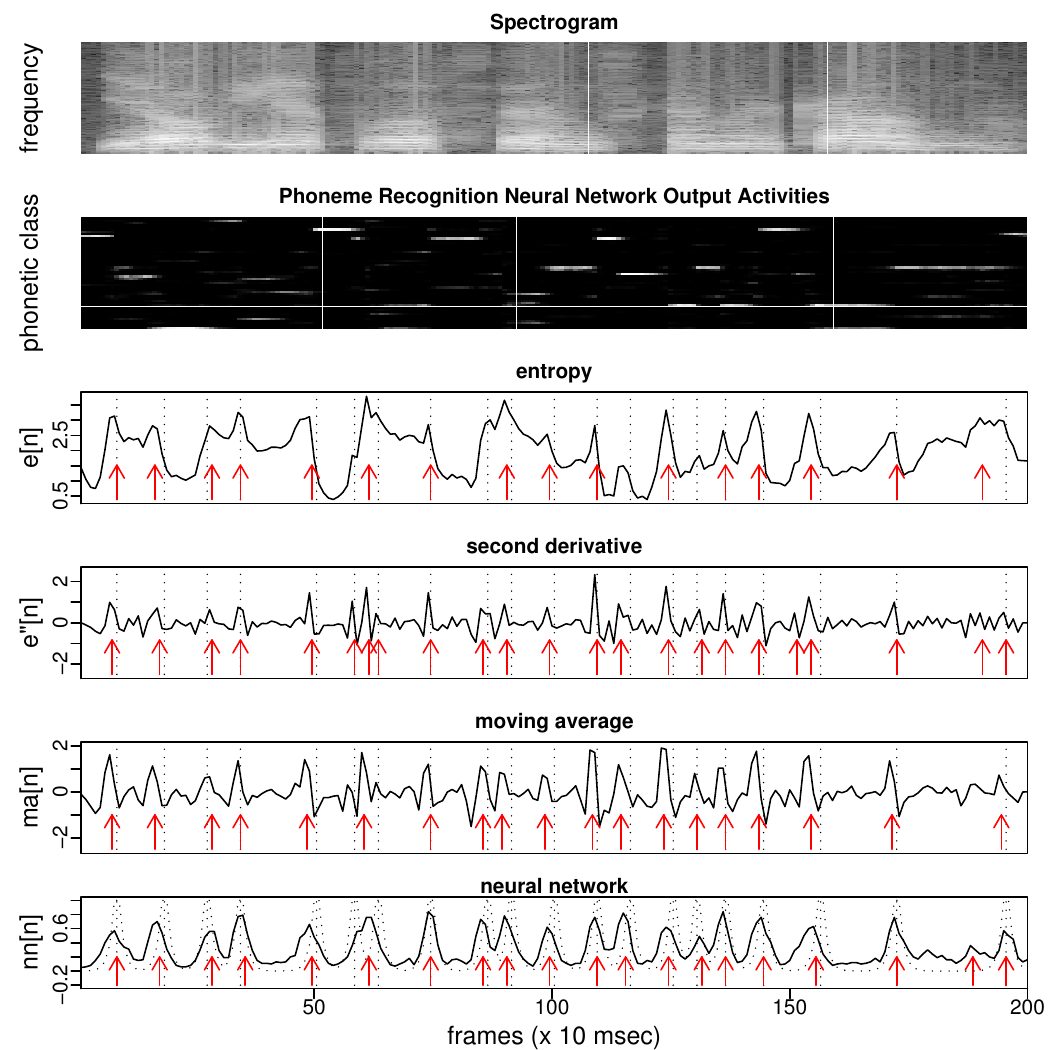}
\caption{Example of test utterance containing the phrase ``jag vet vad
  svenskarna kan'' ([jA:ve:tvA:svenskarnakan]. The last four plots
  show different measures described in the text (continuous line), the
  reference boundaries (dotted lines) and the predicted boundaries
  (arrows). The last plot also include the proximity function (dotted
  line) used as target for the neural network training.} \label{fig:example}
\end{figure}

Figure~\ref{fig:example} illustrates some of the measures defined in
Section~\ref{sec:method} for one test utterance containing the phrase
``jag vet vad svenskarna kan'' ([jA:ve:tvA:svenskarnakan]).

The top plot shows the spectrogram of the signal, the second plot the
output activities $a_i[n]$ of the phoneme recognition network.
The last four plots show four of the measures defined in
Table~\ref{tab:methods}. The vertical dotted lines are the reference
boundaries while the arrows are the predicted boundaries when the
threshold  was chosen in order to obtain a total
number of predicted boundaries closest to the total number of
reference boundaries.

The first measure is the entropy $\mathbf{e[n]}$. The boundary
detection and alignment is acceptable, but a few deletions can be
noticed. The second measure is the second derivative of the entropy
$\mathbf{e^{\prime\prime}[n]}$. This shows an improvement compared to
$\mathbf{e[n]}$ in many cases as the number of deletions is reduced,
however it introduces a number of insertions.

The third measure is the moving average of the second derivative of
the entropy $\mathbf{\mbox{\bf ma}[n]}$. Compared to the second
derivative this eliminates some insertions.

The last measure is the output of the time delayed neural network
$\mathbf{\mbox{\bf nn}(e,e^\prime,e^{\prime\prime},\mbox{\bf
    ma})}$. In this case the bound proximity function
(Equation~\ref{eqn:proximity}) is shown as well as the reference
boundaries in dotted lines. The first observation is that the neural
network does a discrete job in predicting the time evolution of the
proximity function. This in spite of the fact that the extreme values
0 and 1 are never reached. The second observation is that despite one
deletion on the sixth boundary and one insertion before the last, this
measure performs best in the example.

\subsection{Combined Methods: Threshold optimisation}
\begin{figure}
\centering
\begin{tabular}{cc}
\includegraphics[width=0.45\textwidth]{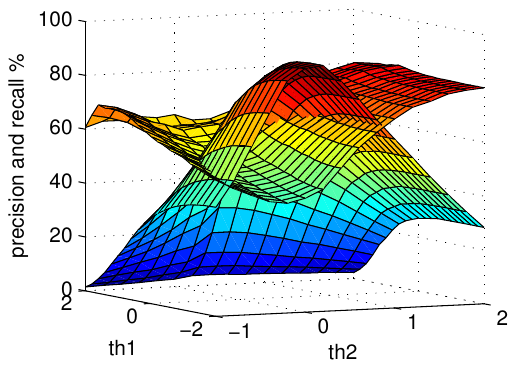} &
\includegraphics[width=0.45\textwidth]{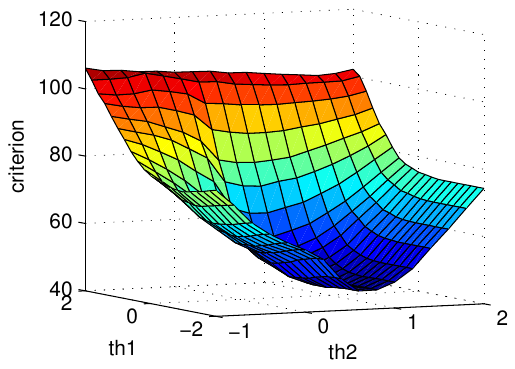} \\
 (a) & (b) \\[5mm]
\includegraphics[width=0.45\textwidth]{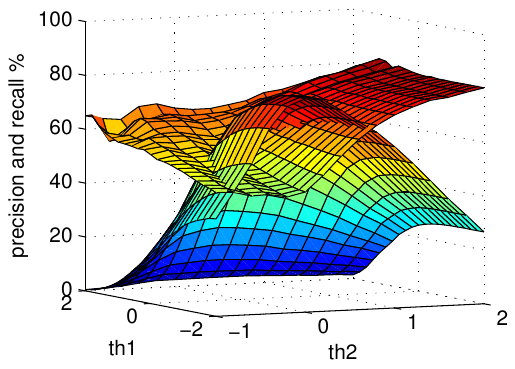} &
\includegraphics[width=0.45\textwidth]{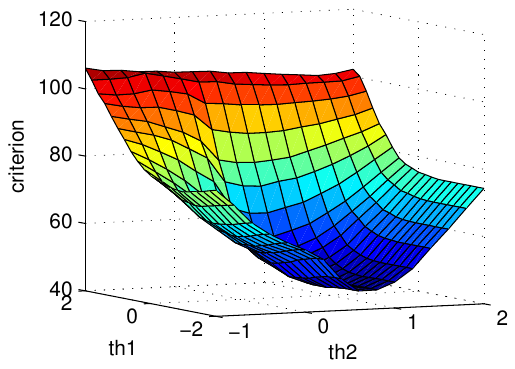} \\
 (c) & (d)
\end{tabular}
\caption{Threshold optimisation for method $\mathbf{e[n]+e^{''}[n]}$
  (a,b) and method $\mathbf{e[n]+\mbox{\bf ma}[n]}$ (c,d). Plots (a)
  and (c) show the recall $R$ (surface with the more pronounced
  maximum) and the precision $P$ as a function of the two thresholds th1
  and th2. The 100\% plane is also shown in the plots. Plots (b) and
  (d) show the optimality criterion crit as a function of th1 and th2.}
 \label{fig:thoptim}
\end{figure}
\begin{figure}
\centering
\includegraphics[width=0.9\textwidth]{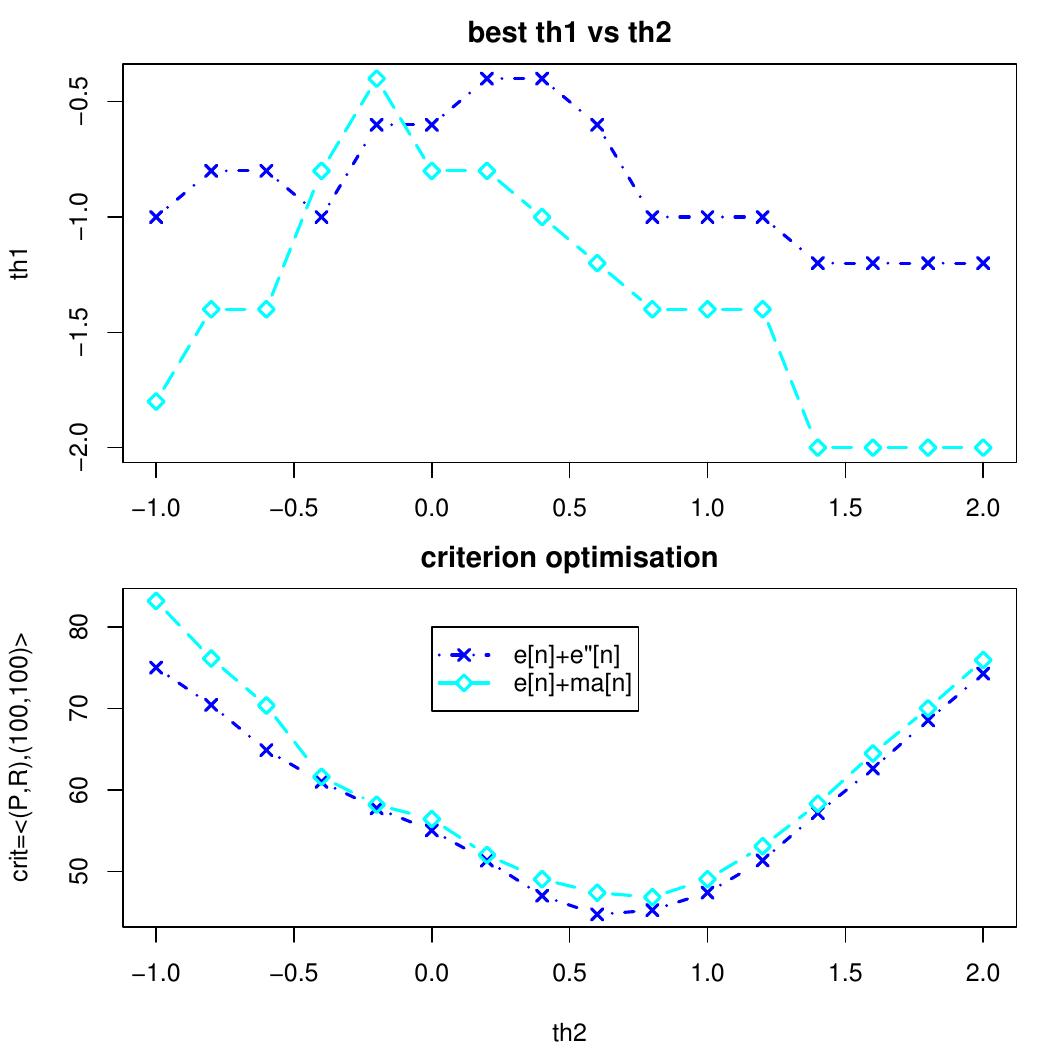}
\caption{Results of the threshold optimisation for methods
  $\mathbf{e[n]+e^{''}[n]}$ and $\mathbf{e[n]+\mbox{\bf
  ma}[n]}$. Above: optimal relative value of th1 in function of
  th2. Below: corresponding value of the criterion in
  Equation~\ref{eq:criterion}}
 \label{fig:thoptim2}
\end{figure}
Figure~\ref{fig:thoptim} shows the optimisation of the two thresholds
th1 and th2 for the two combined methods: $\mathbf{e[n]+e^{''}[n]}$
and $\mathbf{e[n]+\mbox{\bf ma}[n]}$. Precision and recall are shown
on the left plots, while the criterion in Equation~\ref{eq:criterion} is
shown in the right plots. As expected, precision decreases when recall
increases. The precision increases overall when both
thresholds are increased, because the methods become more
selective.  Both the precision and recall surfaces have a bell shape
in the direction of th2 with a minimum (maximum) around th2 $=0$.
This shape is caused by the definition of the relation $\grc$
that picks only one candidate for each region of contiguous measures
that exceed the threshold. Increasing th2 the regions become fewer and
smaller, and, when th2 is higher than the maximum value for the
measure, no region is produced. Decreasing th2 the regions become
larger. Below a certain value of th2 the regions become fewer again
and, for values of th2 lower than the measure's minimum, there
is only one region (and thus one predicted boundary) for each
utterance.

The values of th1 are varied between $-2$ and 2 (in relative
terms). The lower the value of th1, the higher the number of samples
that satisfy the conditions $e[n] > th1$, the more the two combined
methods $\mathbf{e[n]+e^{''}[n]}$ and $\mathbf{e[n]+\mbox{\bf ma}[n]}$
become similar to the simpler methods $\mathbf{e^{''}[n]}$ and
$\mathbf{\mbox{\bf ma}[n]}$.

Figure~\ref{fig:thoptim2} shows the results of the optimisation of the
value of th1 with respect to th2, obtained by minimising the criterion
in Equation~\ref{eq:criterion}. The top plot shows the best value of
th1 as a function of th2 for the two methods $\mathbf{e[n]+e^{''}[n]}$
and $\mathbf{e[n]+\mbox{\bf ma}[n]}$, and the bottom plot shows the
resulting values for the criterion.

\subsection{Result Summary}
\begin{figure}
\centering
\includegraphics[width=0.7\textwidth]{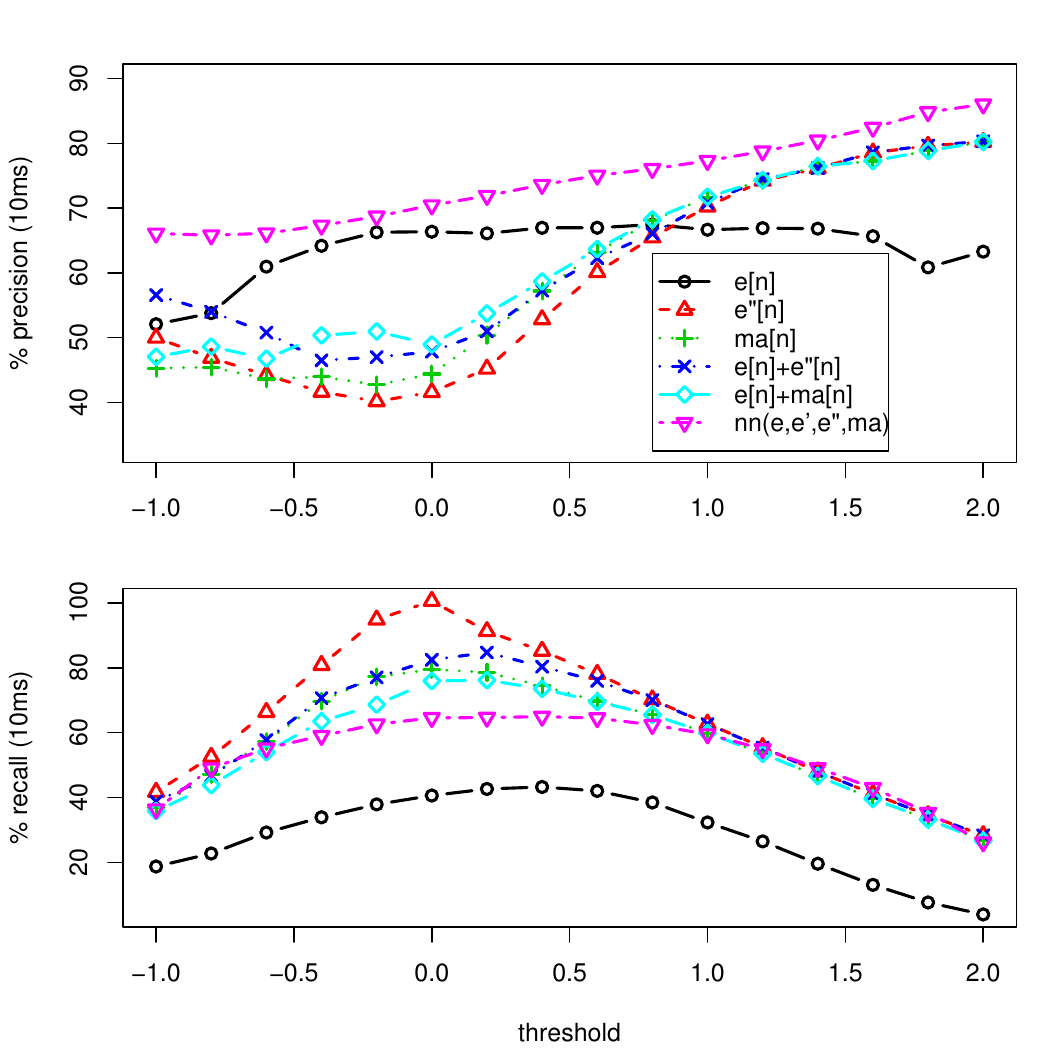}
\includegraphics[width=0.7\textwidth]{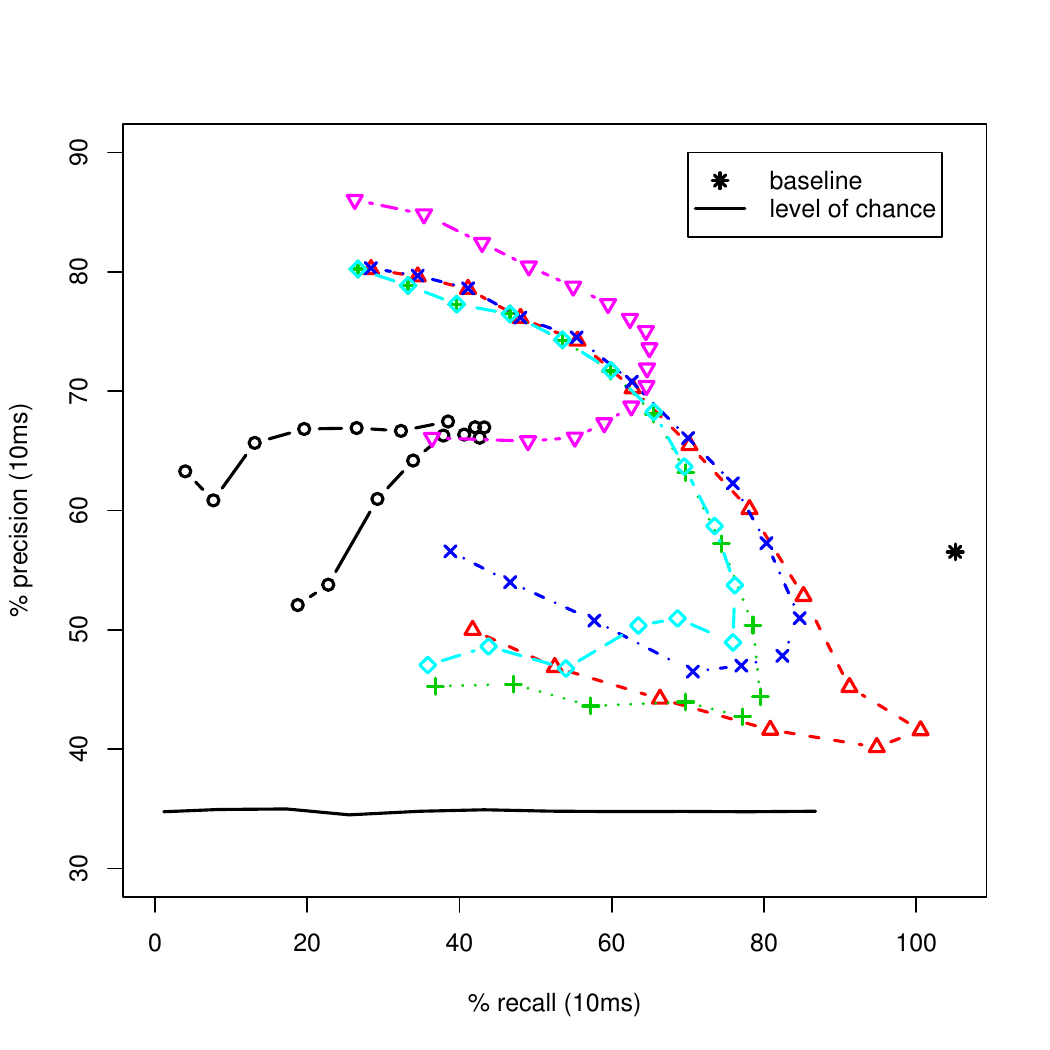}
\caption{Summary of the results for varying relative thresholds. From
  the top: the precision and the recall plotted against the relative
  threshold. The bottom figure plots the precision versus the recall
  when the threshold variable is eliminated. In this plot the baseline
  result and the level of chance, obtained with Monte Carlo simulations,
  are also reported.} \label{fig:resultsummary}
\end{figure}

Figure~\ref{fig:resultsummary} presents a summary of the results
obtained with all methods in Table~\ref{tab:methods} by varying the
value of the relative threshold. In case of combined measures, the
optimal value of the threshold th1, obtained as described in the
previous section, is used for each value of th2. All results in the
figure refer to 10~msec correct boundaries. The first two plots show
the \emph{precision} (10~msec) and the \emph{recall} (10~msec) as
functions of the relative threshold.

In the third plot at the bottom of Figure~\ref{fig:resultsummary}, the
relative threshold variable is eliminated and precision is plotted in
function of recall. The dependency with the relative threshold is
implicitly shown in this plot by the trajectories of each group of
result points. In this last plot the baseline result is reported
as well as the level of chance for different values of recall. The
precision of the level of chance, that was obtained with the Monte
Carlo simulations, is in good agreement with the theoretical value
found in Section~\ref{sec:levelofchance}, with values that range from
34.6\% to 34.9\%. The recall varies between 1\% and 86.8\% when the
total number of predicted boundaries (that is a factor in the
simulations) is varied between 5\% and 250\% of the total number of
reference boundaries. 

The recall for the baseline condition exceeds
100\%. As already noticed in Section~\ref{sec:evaluation}, this is due
to the way the number of correctly detected boundaries is
computed. Given this limitation, the value for the
optimality criterion is unreliable and should not be used to compare
this conditions to the others. The precision is, however, an
interesting figure to determine the performance of this method.

The $\mathbf{e[n]}$ condition shows a low dependency of precision with the
threshold value. However the entropy alone underestimates the number of
boundaries. The best condition with respect to the criterion in
Equation~\ref{eq:criterion} (crit=65.7) gives a precision of 67.0\%
and a recall of 43.3\%.

The $\mathbf{e^{\prime\prime}[n]}$ and $\mathbf{\mbox{\bf ma}[n]}$
conditions clearly outperform $\mathbf{e[n]}$ in terms of the
criterion (45.5 and 47.0 respectively). The first method drastically
improves the recall (78.1\%) at the expense of precision (60.1\%). The
second improves both precision (68.1\%) and recall (65.5\%), even
tough the last to a lesser degree.

The combined measures $\mathbf{e[n]+e^{\prime\prime}[n]}$ and
$\mathbf{e[n]+\mbox{\bf ma}[n]}$, as compared to the corresponding
alone conditions $\mathbf{e^{\prime\prime}[n]}$ and
$\mathbf{\mbox{\bf ma}[n]}$, give in general a higher precision at
the expenses of a lower recall (see the first and second plot in
Figure~\ref{fig:resultsummary}). The performance in terms of the
evaluation criterion is only slightly better than the corresponding alone
conditions with crit=44.8 and crit=46.9 respectively.

Finally the $\mathbf{\mbox{\bf nn}(e,e^\prime,e^{\prime\prime},\mbox{\bf
    ma})}$ condition is best in all cases reaching a precision of 75.0\%
with a recall of 64.5\% and a criterion value of 43.4.



\begin{table}
\caption{Summary of the best results. The values correspond to
    correctness within 10 (20) msec of the reference boundaries,
  $^{*}$) the recall, and therefore the criterion, are in this case
  arbitrary, as the total number of predicted boundaries is a factor
  that does not change the precision. The 100\% recall, and the
  corresponding criterion are reported for comparison with the other
  conditions. $^{**}$) the reason why the recall can be $>$ 100\% is
  explained in the text.}
\label{tab:bestresults}
\centering
\begin{tabular}{lccc}\hline\hline
condition   & precision \% & recall \% & criterion \\ \hline
level of chance & 34.8 (55.0) & 100 (100)$^*$ & 65.2 (45.0)$^*$ \\
baseline        & 56.5 (74.3) & 105 (138)$^{**}$ & 43.8 (46.0) \\
$\mathbf{e[n]}$ & 67.0 (81.0) & 43.3 (52.3) & 65.7 (51.3) \\
$\mathbf{e^{''}[n]}$ & 60.1 (79.8) & 78.1 (85.5) & 45.5 (24.9) \\
$\mathbf{\mbox{\bf ma}[n]}$ & 68.1 (78.5) & 65.5 (86.6) & 47.0 (25.3) \\
$\mathbf{e[n]+e^{''}[n]}$ & 62.3 (77.6) & 75.9 (94.6) & 44.8 (23.0) \\
$\mathbf{e[n]+\mbox{\bf ma}[n]}$ & 68.3 (79.1) & 65.5 (86.3) & 46.9 (25.0) \\
$\mathbf{\mbox{\bf nn}(e,e^{'},e^{''},\mbox{\bf ma})}$ & 75.0 (86.4) &
64.5 (76.2) & 43.4 (27.4) \\
forced alignment & 86.8 (93.9) & 90.8 (98.3) & 16.1 (6.28) \\ \hline\hline
\end{tabular}
\end{table}


Table~\ref{tab:bestresults} summarises the best results obtained with
the different methods. The main results are given using 10~msec limits
of correctness. In the literature about phoneme alignment
\citep[e.g.][]{gs:Hosom2002} the precision is often measured in terms of
percentage of the boundaries falling within 20 msec of the
reference. To allow for comparison the corresponding results are
reported in parenthesis in the table. For the same purpose, the
results obtained by forced alignment as described in
Section~\ref{sec:forcedalign}, are also shown in the table. It is,
however, worth reminding that the task considered in this paper is
different from the phoneme alignment task as no reference
transcription is used in the discussed methods, to find the phonetic
boundaries. The results are therefore not directly comparable.
\section{Discussion}
\label{sec:discussion}
The fundamental limitation of the methods described in this paper is the
fact that the time resolution for the boundary detection is one
frame (10~msec). More precise boundary placements can be achieved with
measures at the speech sample level. This is not a severe limitation
for most speech analysis tasks, also considering that the position of
phonetic boundaries is often ambiguous.

A second limitation is the evaluation criterion. In fact the
computation of number of correctly detected boundaries overlooks
insertions near the boundaries. These can cancel out the number of
deletions of boundaries, giving values of precision and recall that
are misleadingly high. As already discussed in the paper this
problem does not emerge in evaluating the methods here exposed, as
they intrinsically avoid detecting boundaries that are too close in
time. The definition of a distance measure that uses dynamic
programming to guarantee that only one detected boundary is assigned
to each reference boundary, might help estimate with more precision
the amount of insertions and deletions in the results.


The best methods ($\mathbf{e[n]+\mbox{\bf ma}[n]}$ and
$\mathbf{\mbox{\bf nn}(e,e^{'},e^{''},\mbox{\bf ma})}$) seem to
underestimate the number of boundaries in the material. The
corresponding deletions, at a visual inspection, seem to correspond to
local maxima in $e[n]$ that were not correctly detected, giving space
for future improvements. One way to achieve such an improvement is to
implement a more sophisticated decision procedure that could
eliminate the problems discussed in Section~\ref{sec:decision}.

\section{Conclusions} \label{sec:conclusions}
In this paper we propose the use of the entropy of the posterior probabilities
estimated by a neural network for phoneme recognition, as a predictor
of phonetic boundaries.

A number of measures based on time differentiation of the entropy have
been used in isolation and in combination. The best method, based on
a non-linear combination of the entropy and its derivatives achieves
75.0\% precision and 64.5\% recall with 10~msec limits of correctness
and 86.4\% precision and 76.2\% recall with 20~msec limits.
\bibliographystyle{elsart-harv}
\bibliography{IEEEabrv,nolisp2005_3}
\end{document}